\documentclass{appolb}
\usepackage{graphicx}
\usepackage{verbatim}

\begin{document}
\title{Search for eta-mesic helium via
deuteron-deuteron reactions with the
WASA-at-COSY facility
\thanks{Presented at Symposium on Applied Nuclear Physics and Innovative Technologies }%
}
\author{Magdalena Skurzok, Pawe{\l} Moskal, Wojciech Krzemie\'n \\
for the WASA-at-COSY collaboration
\address{M. Smoluchowski Institute of Physics, Jagiellonian University, Cracow, Poland}
\\
}
\maketitle
\begin{abstract}

The $\eta$-mesic nuclei in which the $\eta$ meson is bound with nucleus via strong interaction was postulated already in 1986, however till now no experiment confirmed empirically its existence. The discovery of this new kind of an exotic nuclear matter would be very important for better understanding of the $\eta$ meson structure and its interaction with nucleons. 
The search for $\eta$-mesic helium is carried out with high statistic and high acceptance with the WASA-at-COSY detection setup in the Forschungszentrum J\"ulich.
The search is conducted via the measurement of the excitation function for the chosen decay channels of the $^{4}\hspace{-0.03cm}\mbox{He}$-$\eta$ system. 
Untill now two reactions $dd\rightarrow(^{4}\hspace{-0.03cm}\mbox{He}$-$\eta)_{bs}\rightarrow$ $^{3}\hspace{-0.03cm}\mbox{He} p \pi{}^{-}$ and \mbox{$dd\rightarrow(^{4}\hspace{-0.03cm}\mbox{He}$-$\eta)_{bs}\rightarrow$ $^{3}\hspace{-0.03cm}\mbox{He} n \pi{}^{0}$} have been measured with the beam momentum ramped around the $\eta$ production threshold. This report includes the description of experimental method and status of the analysis.

\end{abstract}

\PACS{PACS numbers come here}
  
\section{Introduction}

\noindent Based on the fact that the interaction between the $\eta$ meson and nucleon
is attractive Haider and Liu postulated the existence of the $\eta$-mesic nuclei~\cite{HaiderLiu1},
in which the neutral $\eta$ meson might be bound with nucleons via the strong
interaction. The existence of $\eta$-mesic nuclei allows to investigate interaction of the $\eta$ meson and the nucleons inside a nuclear matter.~Moreover it would provide information about $\mbox{N}^{*}(1535)$ resonance~\cite{Jido} and about $\eta$ meson properties in nuclear matter~\cite{InoueOset}, as well as about contribution of the flavour singlet component of the quark-gluon wave function of $\eta$ meson~\cite{BassThomas, BassTom}.

According to the theoretical considerations, the formation of the $\eta$-mesic nucleus can only take place if the real part of the $\eta$-nucleus scattering length is negative (attractive interaction), and the magnitude of the real part is larger than the magnitude of the imaginary part~\cite{HaiderLiu2}:

\begin{equation}
|Re(a_{\eta-nucleus})|>|Im(a_{\eta-nucleus})|.\label{eq:eq1}
\end{equation}

\noindent Calculations for hadronic- and photoproduction of the $\eta$ meson gave a wide range of possible values of the s-wave $\eta$N scattering lenght, from $a_{\eta N}$=(0.27 + 0.22i) fm up to $a_{\eta N}$=(1.05 + 0.27i) fm. Such a high values has not exluded the formation of $\eta$-nucleus bound states for a light nuclei as $^{3,4}\hspace{-0.03cm}\mbox{He}$, T~\cite{Wilkin1,WycechGreen} and even for deuteron~\cite{Green}. 
Those bound states have been searched in several past experiments~\cite{Machner,Sokol1,Gillitzer,Berger,Mayer,Pfeiffer}, however none of them gave empirical confirmation of their existence. There are only a promissing experimental signals which might be interpreted as~\mbox{indications} of the $\eta$-mesic nuclei. The ongoing investigations continue at COSY~\cite{KrzeMosSmy,MSkurzok,MoskalSmyrski,Mersmann,Smyrski1,Krzemien1}, JINR~\cite{Anikina}, J-PARC~\cite{Fujioka}, MAMI~\cite{Pheron} and are planned at GSI~\cite{Itahashi}.

\section{Experiment}

\begin{figure}[h]
\centering
\includegraphics[width=8.5cm,height=4.5cm]{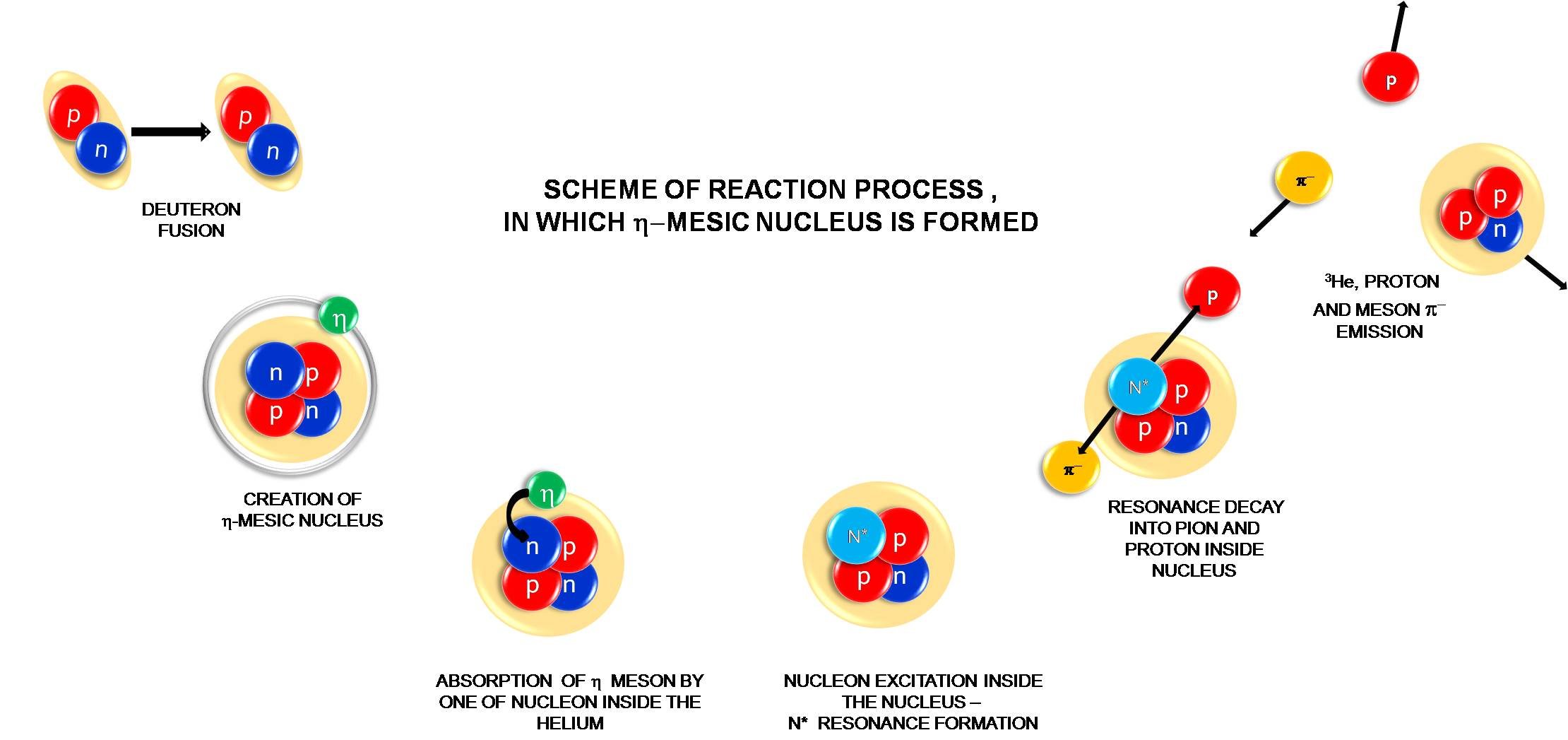}
\caption{Reaction process of the ($^{4}\hspace{-0.03cm}\mbox{He}$-$\eta)_{bs}$ production and decay.}
\label{fig1}
\end{figure}

The measurement of the $^{4}\hspace{-0.03cm}\mbox{He}$-$\eta$ bound states is carried out with unique precision  by means of the WASA detection system, installed at the COSY synchrotron.~Signals of the \mbox{$\eta$-mesic} nuclei are searched for via studying the excitation function of specific decay channels of the \mbox{$^{4}\hspace{-0.03cm}\mbox{He}$-$\eta$} system, formed in deuteron-deuteron collision~\cite{Moskal1,Krzemien}.~The measurement is performed for the beam momenta varying continously around the $\eta$ production threshold.~The~beam ramping technique allows to reduce the systematic uncertainities.~The \mbox{existence} of the bound system should be manifested as a resonance-like structure in the excitation curve of eg. $dd\rightarrow(^{4}\hspace{-0.03cm}\mbox{He}$-$\eta)_{bs}\rightarrow$ $^{3}\hspace{-0.03cm}\mbox{He} p \pi{}^{-}$ reaction
below the $dd\rightarrow$ $^{4}\hspace{-0.03cm}\mbox{He}$-$\eta$ reaction threshold. 
The reaction kinematics is schematically presented in Fig.~\ref{fig1}.


The deuteron - deuteron collision leads to the creation of the $^{4}\hspace{-0.03cm}\mbox{He}$-$\eta$ bound system. The $\eta$ meson can be absorbed by one of the nucleons inside helium and may propagate in the nucleus via consecutive excitation of nucleons to the $\mbox{N}^{*}(1525)$ state~\cite{Sokol} until the resonance decays into the pion-nucleon (n, $\pi^{0}$ or p, $\pi^{-}$) pair outgoing from the nucleus~\cite{KrzeMosSmy}. The relative nucleon-pion angle is equal to $180^\circ$ in the  $\mbox{N}^{*}$ reference frame and it is smeared by about $30^\circ$ in the center-of-mass frame due to the Fermi motion of the nucleons inside the helium nucleus.

In June 2008 a search for the $^{4}\hspace{-0.03cm}\mbox{He}$-$\eta$ bound state was based on the measurement of the excitation function of the $dd\rightarrow$ $^{3}\hspace{-0.03cm}\mbox{He} p \pi{}^{-}$ reaction near the $\eta$ production threshold.~During the experiment the deuteron beam momentum was varied continuously from 2.185~GeV/c to 2.400~GeV/c which corresponds the excess energy variation from -51.4 MeV to 22 MeV. 
Excitation function was determined after applying cuts on the p and $\pi^{-}$ kinetic energy distribution and the $p - \pi^{-}$ opening angle in the CM system~\cite{Krzemien_PhD}. The relative normalization of points of the $dd\rightarrow$ $^{3}\hspace{-0.03cm}\mbox{He} p \pi{}^{-}$ excitation function was based on the quasi-elastic proton-proton scattering.
In the excitation function there is no structure which could be interpreted as a resonance originating from decay of the $\eta$-mesic $^{4}\hspace{-0.03cm}\mbox{He}$~\cite{Krzemien_MESON2012,Wojtek_publ}.

During the second experiment, in November 2010, two channels of the $\eta$-mesic helium decay were measured:  $dd\rightarrow(^{4}\hspace{-0.03cm}\mbox{He}$-$\eta)_{bs}\rightarrow$ $^{3}\hspace{-0.03cm}\mbox{He} p \pi{}^{-}$ and  $dd\rightarrow(^{4}\hspace{-0.03cm}\mbox{He}$-$\eta)_{bs}\rightarrow$ $^{3}\hspace{-0.03cm}\mbox{He} n \pi{}^{0} \rightarrow$ $^{3}\hspace{-0.03cm}\mbox{He} n \gamma \gamma$~\cite{MSkurzok}. The measurement was performed with the beam momentum ramping from 2.127 GeV/c to 2.422 GeV/c, corresponding to the range of the excess energy \mbox{Q$\in$(-70,30)~MeV}. Until now the \mbox{$dd\rightarrow(^{4}\hspace{-0.03cm}\mbox{He}$-$\eta)_{bound}\rightarrow$ $^{3}\hspace{-0.03cm}\mbox{He} n \pi{}^{0} \rightarrow$ $^{3}\hspace{-0.03cm}\mbox{He} n \gamma \gamma$} reaction is analysed. The $^{3}\hspace{-0.03cm}\mbox{He}$ is identified in the Forward Detector based on the $\Delta E$-$E$ method. The neutral pion $\pi^{0}$ is reconstructed in the Central Detector from the invariant mass of two gamma quanta originating from its decay while the neutron four-momentum is calculated using the missing mass technique.

The excitation function for the $dd\rightarrow$ $^{3}\hspace{-0.03cm}\mbox{He} n \pi{}^{0} \rightarrow$ $^{3}\hspace{-0.03cm}\mbox{He} n \gamma \gamma$ reaction is studied for the "signal-rich" region corresponding to the momenta of the $^{3}\hspace{-0.03cm}\mbox{He}$ in the CM system below 0.3 GeV/c and the "signal-poor" region for the $^{3}\hspace{-0.03cm}\mbox{He}$ CM momenta above 0.3 GeV/c. The contributions from different background reactions is under investigation. 

\section{Acknowledgements}
\noindent We acknowledge support by the Foundation for Polish Science - MPD program, co-financed by the European
Union within the European Regional Development Fund, by the Polish National Science Center through grant No. 2011/01/B/ST2/00431 and by the FFE grants of the Research Center Juelich.



\newpage 
\begin{thebibliography}{}


\bibitem{HaiderLiu1} Q. Haider, L. C. Liu, Phys. Lett. \textbf{B 172}, (1986) 257.

\bibitem{InoueOset} T. Inoue, E. Oset, Nucl. Phys. \textbf{A 710}, (2002) 354.

\bibitem{Jido} D. Jido, H. Nagahiro, S. Hirenzaki, Phys. Rev. \textbf{C 66}, (2002) 045202.

\bibitem{BassThomas} S. D. Bass, A. W. Thomas, Phys. Lett. \textbf{B 634}, (2006) 368.

\bibitem{BassTom} S. D. Bass, A. W. Thomas, Acta Phys. Pol. \textbf{B 41}, (2010) 2239.

\bibitem{HaiderLiu2} Q.~Haider, L. C.~Liu, Phys. Rev. \textbf{C 66}, (2002) 045208.

\bibitem{Wilkin1} C. Wilkin, Phys. Rev. \textbf{C 47}, (1993) R938.

\bibitem{WycechGreen} S.~Wycech, A. M. Green and J. A. Niskanen, Phys. Rev. \textbf{C 52}, (1995) 544.

\bibitem{Green} A. M. Green, Phys. Rev. \textbf{C 54}, (1996) 1970.

\bibitem{Machner} H. Machner, Acta Phys. Polon. \textbf{B 41}, (2010) 2221.

\bibitem{Sokol1} G. A. Sokol et al., arXiv:nucl-ex/0012010 (2000).

\bibitem{Gillitzer} A. Gillitzer, Acta. Phys. Slovaca \textbf{56}, (2006) 269.

\bibitem{Berger} J.~Berger et al., Phys. Rev. Lett. \textbf{61}, (1988) 919.

\bibitem{Mayer} B.~Mayer et al., Phys. Rev. \textbf{C 53}, (1996) 2068.

\bibitem{Pfeiffer} M. Pfeiffer et al., Phys. Rev. Lett. \textbf{B 92}, (2004) 252001.

\bibitem{KrzeMosSmy} W.~Krzemie\'n, P.~Moskal, J.~Smyrski, Acta Phys. Polon. Supp. \textbf{2}, (2009) 141.

\bibitem{MSkurzok} M. Skurzok, P. Moskal, W. Krzemien, Prog. Part. Nucl. Phys. \textbf{67}, (2012) 445.

\bibitem{MoskalSmyrski} P. Moskal, J. Smyrski, Acta Phys. Polon. \textbf{B 41}, (2010) 2281.

\bibitem{Mersmann} T.~Mersmann et al., Phys. Rev. Lett. \textbf{98}, (2007) 242301.

\bibitem{Smyrski1} J.~Smyrski et al., Phys. Lett. \textbf{B 649}, (2007) 258.


\bibitem{Krzemien1} W. Krzemie\'n et al., Int. J. Mod. Phys. \textbf{A 24}, (2009) 576.



\bibitem{Anikina} M. Kh. Anikina et al., arXiv:nucl-ex/0412036 (2004).

\bibitem{Fujioka} H. Fujioka, K. Itahashi, Acta. Phys. Pol. \textbf{B 41}, (2010) 2261.

\bibitem{Pheron} F. Pheron, et al. Phys. Lett. \textbf{B709}, (2012) 21.

\bibitem{Itahashi} K. Itahashi, et al. Prog. Theor. Phys.  \textbf{128}, (2012) 601.


\bibitem{Moskal1} P.~Moskal et al., Int. J. Mod. Phys. \textbf{A 22}, (2007) 305.
 
\bibitem{Krzemien} W. Krzemie\'n, arXiv:nucl-ex/1101.3103, (2011). 

\bibitem{Sokol} G. A. Sokol et al., arXiv:nucl-ex/0106005 (2001).

\bibitem{Krzemien_PhD} W.~Krzemie\'n, \textit{PhD Thesis, Jagiellonian University}, arXiv:nucl-ex/1202.5794 (2011).


\bibitem{Krzemien_MESON2012} W.~Krzemie\'n, MESON 2012 proceedings (2012).

\bibitem{Wojtek_publ} P. Adlarson et al., Phys. Rev. \textbf{C 87}, (2013) 035204.

\end{thebibliography}
\end{document}